# Segmentation, Incentives and Privacy

Kobbi Nissim[*]     Rann Smorodinsky[†]     Moshe Tennenholtz[‡]

July 17, 2017

**Abstract**

Data driven segmentation is the powerhouse behind the success of online advertising. Various underlying challenges for successful segmentation have been studied by the academic community, with one notable exception − consumers incentives have been typically ignored. This lacuna is troubling as consumers have much control over the data being collected. Missing or manipulated data could lead to inferior segmentation. The current work proposes a model of prior-free segmentation, inspired by models of facility location, and to the best of our knowledge provides the first segmentation mechanism that addresses incentive compatibility, efficient market segmentation and privacy in the absence of a common prior.

---

[*]Department of Computer Science, Georgetown University. Work partly done while the author was in Ben-Gurion University and the Center for Research on Computation and Society, Harvard University. Supported by NSF grant CNS-1237235, grants from the Sloan Foundation, and ISF grant 276/12. kobbi.nissim@georgetown.edu.

[†]Faculty of Industrial Engineering and Management, Technion – Israel Institute of Technology, Haifa 32000, Israel. This work was generously supported by the Israeli Science Foundation (grant No. 276/12), The Technion-Microsoft EC research center, Technion VPR grants and the Bernard M. Gordon Center for Systems Engineering at the Technion. rann@ie.technion.ac.il.

[‡]Faculty of Industrial Engineering and Management, Technion – Israel Institute of Technology, Haifa 32000, Israel. moshet@ie.technion.ac.il. The support of EU project 740435 - MDDS is gratefully acknowledged.

# 1  Introduction

Segmentation is considered as one of the crucial marketing processes for many firms and is viewed as the process of dividing the market into groups of customers or consumers with similar needs.[1] Segmentation is the core process leading to, inter alia, market positioning, product definition, campaigns, pricing and distribution channels.[2] [3]

Market segmentation has traditionally been based on geographic and demographic attributes of the consumers. However, with the prevalence of consumers' online footprint and in particular with usage and consumption data many firms find psychographic (also known as 'lifestyle') and behavioral segmentation much more efficient (see [29]). One can safely say that segmentation, and in particular psychographic and behavioral segmentation, is central in multi billion dollar markets. The online advertising business alone, which is primarily based on behavioral segmentation (sometimes referred to as *targeting*), is currently estimated at nearly 50 USD billion annually in the USA alone, with 17% growth rates [26].

Modern behavioral and psychographic segmentation techniques, apart from relying on large amounts of data, also differ from the more traditional segmentation strategies in that they utilize diverse data sources, most of which are on-line. Some examples of these diverse data sources are small files called *cookies* that are downloaded to the user's browser and facilitate the collection of browsing history, search engines that collect the history of search terms, on-line retailers (e.g., eBay or Amazon) that collect purchase records, and services where subscription and registration is required that collect our personal information (such as age, address, gender). All these data sources can now be integrated into the segmentation process. One important thing to note about all these data sources is that the individual consumer can have significant control on the data provided. For example, he may choose not to install cookies or to provide partial, or even false information in a registration process. He may choose not to buy things or restrict his shopping to some specific category. He may choose to be a proactive member of a social network (e.g., use Facebook's *like* functionality and post to his wall frequently) or use it in a passive manner only. All of this stands in stark contrast with the more traditional segmentation. Indeed geographic and demographic segmentation relied mostly on *publicly available* data and hence was not subjected to potential manipulation by consumers.

---

[1]The Wikipedia entry for 'Market segmentation' begins as follows: "Market segmentation is a marketing strategy that involves dividing a broad target market into subsets of consumers who have common needs and priorities, and then designing and implementing strategies to target them. Market segmentation strategies may be used to identify the target customers, and provide supporting data for positioning to achieve a marketing plan objective. Businesses may develop product differentiation strategies, or an undifferentiated approach, involving specific products or product lines depending on the specific demand and attributes of the target segment."

[2]For more on segmentation and (third degree) price discrimination see Varian [45] for the the monopoly setting and Stole [42]) for the oligopoly setting.

[3]To support this consider the following quote from Wind [48]: *"Marketing segmentation long has been considered one of the most fundamental concepts of modern marketing. In the 20 years since the pioneering works by Wendell Smith, segmentation has become a dominant concept in marketing literature and practice. Besides being one of the major ways of operationalizing the marketing concept, segmentation provides guidelines for a firm's marketing strategy and resource allocation among markets and products. Faced with heterogeneous markets, a firm following a market segmentation strategy can increase the expected profitability . . . realizing the potential benefits of market segmentation requires both management acceptance of the concept and an empirical segmentation study . . ."*

Surprisingly, most firms involved in segmentation take it for granted that the data they collect indeed accurately reflects the consumer and so, for the purpose of segmentation, they choose to ignore the possible strategic manipulation of the data. In the jargon of classical mechanism design consumers are assumed truthful, even if this cannot be justified either theoretically or empirically. In this paper we take issue with the (implicit) assumption of truthfulness on consumers' behalf and study segmentation under the working hypothesis that players do act strategically. Thus, we study segmentation when consumers are aware that their data is collected and used for segmentation. To establish a coherent model we actually assume that consumers are fully rational and also knowledgeable about the way data is used to further encourage them to consume.

The model we propose is inspired by the following stylized description of the segmentation process as a four stage process:

1. Consumers are asked to provide the data that describes them. One can think of this as a vector of inputs, where each entry is the value for some given attribute (e.g., gender, income, frequency of visits to Facebook, total dollar consumption on eBay, etc.). In our abstract model we refer to this as a point in the $N$-dimensional unit cube, $T = [0, 1]^N$. Whereas the true data is some point $t \in T$ the consumer may report any value $b \in T$, as the data is not verifiable.

2. The firm partitions the set of consumers into a fixed number of segments, say $K$ (in our model $K$ will be set exogenously). Informally, segments are generated such that each one is as homogeneous as possible and any two segments are as distinct as possible. In our model the notion of homogeneity and distinction are captured via a metric on $T$.

3. Within each segment a *representative consumer* is identified. This may be an actual consumer or a virtual one. The representative consumer is the one that maximizes the average similarity with the consumers within the segment. In some sense this step and the previous one may be flipped and at the preliminary stage $K$ representative agents are identified in such a way that minimizes the average distinction between a consumer and the representative agent most similar to him. The choice of $K$ representative agents naturally induces a partition of the set of consumers into $K$ segments. Indeed, this reversed process is the one captured by our model and mechanism.

4. The firm's endeavors (products, campaigns, pricing, etc.) are focused to satisfy the $K$ representative consumers.

To demonstrate the above process consider an on-line retailer such as Booking.com which has at its disposal a variety of promotional tools and would like to use the optimal combination of such tools for its registered user base. For example, Booking.com may propose discounts at countryside boutique hotels or weekend getaways in big cities, it may offer a buy-two-nights-get-one-free for early booking or last-minute all inclusive deals and so on and so forth. Proposing the full arsenal of promotions to all of its customers may prove unproductive and possibly not optimal. The typical way to better match propositions with customers is via segmentation. To identify these segments Booking.com would refer to some customer data. Such data may be collected from internal sources (e.g., the registration form or past reservations or bookings on Booking.com) or from external sources such as purchasing data from social networks, data from cookies in some on-line retail stores and search queries to name a few. Once the data is collected the customer base can be segmented.

The data driven segmentation process may lead to a segment defined, for example, by activity on children entertainment websites (e.g., www.nick.com, the Nickelodeon website) coupled with the age between 25-40 (taken from the registration form), suggesting a family with young kids. Alternatively, activity on Facebook coupled with a recent purchase from online jewelry retailers may determine an 'upper-class newly weds' segment. Once the segmentation process is over it is the role of the marketing department to depict the 'representative customer' to which the optimal menu of propositions is tailored.[4]

The challenges underlying this process are overwhelming. The need to collect and save large amounts of data, the need to make cross references between a variety of data sources, the optimal way to save the data in a concise and aggregated manner without losing too much knowledge, the complexity of computing optimal segments and more. Many of these challenges have been studied and (partly) resolved. However, one aspect of the segmentation process which has remained untreated (to the best of our knowledge) is the issue of truthfulness of the data, or more generally the underlying incentives of the consumers and how they affect the outcome of the segmentation process. In fact, the process of behavioral segmentation, which is key to successful marketing for many firms, is a natural mechanism design challenge. It is quite surprising to witness, therefore, the lack of literature on segmentation within the mechanism design community. This motivates the current study, where we frame segmentation as a mechanism design challenge and propose a mechanism that is nearly optimal while accounting for consumers' incentives.

## 1.1 Targeted advertising

Our segmentation model is directly inspired by the 'targeted advertising' industry, which is one of the fastest growing industries in the recent decade. This is evident from the current market size. In USA alone the estimated revenues for Internet advertising in $2014$, of which targeted advertising forms the lion's share, neared $\$50$ billion, up $17\%$ from the the $2013$ revenues (see [26]).

---

[4]In many retail companies the process of defining the 'representative customer' is quite an intricate and complex process. They are often given a full background, including age, gender, names, profession, habits and so on. This is then used as a means to align all the marketing efforts - for example, pricing of goods, communications channels, packaging and promotions.

Targeted advertising is primarily centered around behavioral segmentation for on-line advertising. To demonstrate targeted advertising we consider a firm, say Nike, which would like to maximize the consumers attention to its new product (e.g., new line of running shoes) via its on-line advertisement. To do so the firm plans to post advertisements on the Sports Illustrated website. The firm may design a few different variants of the advertisement, say with different colors, themes and messages. The underlying process which determines the nature of each of these advertisements is segmentation. Once the segmentation process is done ad variants are designed to optimally appeal to each segment. In real-time, that is when a consumer visits the aforementioned publisher's website, the firm must decide (almost instantaneously) which ad variant to expose to the consumer with the objective that the consumer will click on the ad which, in turn, may lead to purchasing the firm's product. Thus, the challenge for targeted advertising is choosing a set of advertisements and associating with each consumer the one most likely to get his attention with the overall objective of maximizing the click through rate (CTR) on advertisements.

## 1.2 Our Contribution

This paper initiates the study of consumer segmentation schemes and the way consumers' incentive constraints effect the quality and quantity of the data consumers disclose. This data, in turn, effects the quality of the resulting segments. The context of our model is that of on-line advertising and hence we consider the standard objective function used in this industry which is Click Thru Rates (CTR). We provide a rigorous model for the segmentation design challenge and present a segmentation mechanism that is incentive compatible and (almost) optimal.[5] The scheme we provide uses a biased coin to decide between a quantal response mechanism (one which assigns higher probabilities to better segmentations) and an oblivious mechanism which mostly ignores the data. The randomness at the heart of our construction is powerful enough to induce truthfulness and (almost) efficiency even for a strong solution concept such as a single elimination of dominated strategies.

Our solution concept, which is adequate for the prior free setting we study, is that of a strategy tuples which survive a single deletion of dominated strategies.

In addition, we discuss privacy issues related to the usage of on-line consumer data and show that, in addition, our segmentation process is relatively immune to privacy concerns.

## 1.3 Related Work

We turn to discuss related work from a variety of disciplines: economics, marketing, operations research and machine learning.

---

[5]The notion of incentive compatibility is ambiguous as it depends on which solution concept one has in mind. The solution concept considered in this paper is that of undominated strategies.

The economics literature deals extensively with the notion of differential pricing (see e.g. [45] and [42] for some overview). In that setting different product prices may be assigned to different segments of the population in various ways. In most of the related papers the segment structure – the (typically finite) set of types and the prior distribution over this set, is known in advance. The seller and consumers then compete for the informational rents. The crux of the seller's challenge is to minimize the informational rents of consumers by providing a price menu such that each type self selects to a price in a way that maximizes profits. In particular, from the outset the firm knows the finite set of types and the proportion of the population associated with each type. Hence, deciding on the segments is superfluous. An exception is the work by [3] where the authors consider a model of asymmetric information on consumers' willingness to pay for the good (see also [46] for further discussion along these lines). In their two stage model, consumers are offered a single price for the good which they can accept or reject. In a second stage differential prices are offered based on the reaction to the initial price (as recorded by a cookie in an online shopping setting). Thus, in the first stage the seller collects information in addition to generating revenues. This work considers a finite and single dimensional type model (two types actually) and a common prior. In particular the common prior assumption of [3] implies that some study on the population has already been conducted in some a-priori stage (which they do not model). More generally, the common prior assumption is not well-suited for studying primal settings where nothing is known.

In contrast, we focus on a prior-free type space. This captures a situation where (almost) nothing is known about the consumers, and the segmentation process is done over a clean slate. In addition, we assume that the number of possible types is substantially larger than the number of consumers and definitely larger than the number of segments the seller can handle. This assumption also allows us to model primal settings where nothing is known about the consumers. It also complicates the challenge. To demonstrate why this makes the segmentation challenge more complex consider the complete information version of the two models, ours and that in [3]. In [3], the number of prices equals the number of types which is just the number of segments. Thus, each segment is offered his maximal willingness to pay. In contrast, even with complete information, the seller is left with the complex task of optimal segmentation (or clustering as often termed in other disciplines).

On another aspect, our segmentation challenge is simpler. Whereas the differential pricing models put the consumer and seller with opposite interests, our model focuses on some abstract fitness between goods and segments and so interests are much more aligned.

The economics literature on marketing strategies (beyond pricing) is somewhat limited. One such work is Johnson and Myatt [27]. They provide a framework for analyzing transformations of demand and use it to study a monopolist's optimal advertising and marketing strategy. They consider a monopolist with either a single product or a a given set of differentiated product variants.

The theoretical literature in economics that studies segmentation and related product differentiation dates back to the seminal paper of Harold Hotelling [24]. In Hotelling's setting the firms compete on location and each firm wants to maximize the size of the segment of consumers which it attracts. Hotelling and others show that in some cases the competition results in product homogeneity while under other assumptions significant product differentiation is observed. In this work, as well in a large body of follow up literature on 'spatial economics', the consumers hold no private information and do not play a strategic role, whereas the emphasis is on the strategic interaction among firms and how this is related to profitability. In particular, almost no attention was given in the body of work on spatial economics to consumers' overall welfare. One exception is Lancaster [30] who studies the social welfare implications of product differentiation instead of the firms' profitability, which drive preceding literature.

More recently, the question of optimal multiple facility location was studied from the point of view of a social planner who can dictate the location of such facilities. In this mechanism design challenge it is the consumers who have private information and do play a strategic role. A central planner asks agents (consumers) to report their location (which is private information) and then dictates the location of a set of $K$ facilities. The objective function of the planer is to minimize the distance agents have to travel to the nearest facility. To recast the facility location to the segmentation jargon one can think of the set players that utilize a certain facility as segment. Therefore, determining the $K$ locations for the facilities induces a segmentation, where the facilities serve as 'centroids'. Note that in the facility location framework it is assumed that once facilities are located each agent may decide to use any of the $K$ facilities and not necessarily the one closest to the position he primarily announced. In contrast, the segmentation challenge allows for the planner to dictate not only the location of the 'centroids' but also which centroid —(or segment) each consumer is associated with. This turns out to be a significant component for obtaining positive results. In fact, work on multiple facility location turned out to be challenging [31, 19], and despite the interesting results obtained, these could not address the challenge of (approximately) optimizing expected social welfare. In contrast, the segmentation challenge is surmountable as we shall demonstrate. A preliminary discussion of the facility location in the context of such a dictatorial planner appears in our earlier work [35], which only discusses a 'linear' setting with a finite state space.

Segmentation and incentives have already been studied in the past in the context of *Hedonic games*. In such games players have a preference relation over the set of coalitions they would like to be part of. A segmentation is Nash-stable if no player would like to deviate from its current segment to another one. Players' underlying preference on segments may be arbitrary or may be derived from some proximity (similarity) between players, as in our context. Additional stability notions such as core stability have also been studied (see [6, 9, 10], among others). While the spirit of our model has some similarity with the above, our model and approach are vastly different. Central to our motivation and model is the role of the mechanism designer who has his own objective function, independent of players' preferences. Such a designer is not part of the hedonic game model.[6] Another distinction between the literature on hedonic games and our work is that the former was mostly studied in the context of complete information whereas we study segmentation with incomplete information.

---

[6]Although these games have not been studied in the context of mechanism design the social welfare implications of equilibrium have been. In particular [18] study the price of anarchy in such games.

Market segmentation is also a central topic in within the disciplines of marketing, machine learning and optimization. However, in this body of work the consumer either has no private information or, if he does, is assumed to act in a non-strategic way and truthfully discloses his private information.

The first academic treatment of market segmentation, to the best of out knowledge, took place in the middle of 1950s by Wendell Smith [41]. In his work Smith advocates market segmentation being an important tool to enable marketers to better meet customer needs. He views it as development of the demand side of the market that represents a rational and more precise adjustment of product and marketing effort to consumer or user. Much of the literature that evolved from Smith's original work was about the parameters and families of parameters according to which segmentation should be done. In fact, by and large, four segmentation bases have been recognized throughout the years in the marketing literature (see [29]): geographic segmentation, demographic segmentation, psychographic segmentation, and behavioral segmentation. The segmentation base chosen to subdivide a market will depend on many factors such as the type of product, the nature of demand, the method of distribution, the media available for market communication, and buyers' motivation [13]. The marketing methods discussed in this literature, and in particular segmentation methods (see a discussion of methods of segmentation in [20]), almost entirely ignore game-theoretic considerations on the part of consumers. Thus, the available data underlying segmentation is (implicitly) considered truthful. Specific major challenges tackled by market segmentation research are the identification of variables that are crucial for segmentation (e.g., is socio-economic status a good attribute? [7]), in addition to statistical approaches (e.g., [5]) and the development of general methodologies for market segmentation as part of the "marketing mix" (see, e.g., [47]).

In contrast, the optimization and machine learning communities, have focused their efforts on understanding the algorithmic and statistical challenges underlying segmentation. The segmentation challenge is known as the $k$-median problem in the theory of optimization and as the study of 'centroids' in the Machine Learning community. A large set of $n$ points in an arbitrary metric space is given. The challenge is to efficiently identify a subset of $k$ points in a way that the sum of distances of points from their closest centroid, respectively, is minimized (see, e.g., [12]). A related problem is the $k$-center problem in which the aim is to minimize the maximal distance over all distances from the selected centroids (see, e.g., [23]). Notice that such a selection of centroids (plus some tie-breaking rule) naturally induces a partitioning/segmentation of the points.

In the above literature it is taken for granted that the underlying data set is truthful and has not been manipulated. Thus, consumers incentives and possible strategic behavior is ignored. Hypothetically, were such incentives accounted for, many of the results and algorithmic achievements could fail.

The current paper falls within the larger framework of approximate mechanism design without monetary transfers initiated by [37]. From a technical perspective we adopt and generalize on the ideas presented in our earlier work [35, 34] where we introduce approximate mechanism design without money in a setting where the set of types as well as the set of social alternatives is finite, as opposed to the current work. In the finite case we were able to leverage the natural lower bound over the difference between a user's utility for two distinct outcomes. This lower bound does not hold when the set of outcomes is a continuum, as we naturally have in our segmentation problem. Instead, we design an oblivious mechanism that is "strongly incentive compatible" in the sense that the loss from misreporting depends quadratically in how much the user misreports. As a result, the incentive compatibility notion we work with is that of undominated strategies, as opposed to the more prevalent one of ex-post Nash equilibrium strategies used in [35]. Obviously the set of strategy tuples we consider in the current work (those surviving a single elimination stage of dominated strategies) contains the set of all strategy tuples that form an equilibrium. Hence, implementation in undominated strategies is more robust. Note, in particular, it relies on players' rationality but does does not rely on players knowing others are rational. The reason we can push the techniques further is due to the focus on a narrow design problem, which we argue is economically important, as opposed to our previous work which addressed more abstract settings.

Our earlier work [35, 34] leveraged the concept of differential privacy [16], and the recently established connection between differential privacy and mechanism design. Differential privacy is a definition of privacy that has emerged in the computer-science literature on foundations of data analysis. In a differentially private mechanism, every agent's influence on the outcome distribution is bounded in the sense that by changing its input the agent can influence the probability of each outcome by at most a factor of $1 + \epsilon$. McSherry and Talwar proposed in their seminal work [33] that differential privacy can be used as a tool for mechanism design, a proposition that was further developed in our work [35].[7] One particular paper worth mentioning is [14] who study a two stage model where the consumption behavior in the first stage effects the segment the consumer is associated with in the second stage, where the segmentation is used for targeted advertising. The paper then investigates the effect of the level of privacy protection (manifested in the privacy parameter of differential privacy ) chosen at stage one (which is thought of as the result of some privacy regulation) over the equilibrium behavior. The authors show that introducing privacy leads to non intuitive phenomena. For example, a higher level of privacy can entail more information disclosure on the consumer's type.

Last but not least, our work adds to the body of work on *virtual implementation*, and in particular robust virtual implementation. In many settings implementation is known to be impossible and so researchers have suggested less challenging, yet meaningful replacements to implementation. One such avenue of research is that of virtual implementation, where the challenge of implementing a function is replaced with implementing an $\epsilon$-approximation of that function, as we do here. This line of work initially focused on implementation in complete information environments (see Matsushima [32] and Abreu and Sen [2]), later extended to private information settings (e.g., to Abreu and Matsushima [1], Duggan [15], and Serrano and Vohra [39, 40] ) and more recently to private information settings without a common prior as we do here (see Bergemann and Morris [8]).

---

[7]For more on the connection between privacy and mechanism design we suggest the review by Heffetz and Ligett [22], a survey by Pai and Roth [36].

# 2 Model

Let $\mathcal{N}$ be a set of $N$ consumers where each consumer is associated with a point in the $I$ dimensional cube, $T = [0,1]^I$. Thus, consumer $n$ is described by the vector $t^n \in T$ of his attributes. The vector $t^n$ is the $n$-th consumer's private information.

A firm would like to advertise its products on-line to this audience in such a way that consumers observing the ad will click on it and would be re-directed to the firm's on-line shop. Thus, the firm would like to generate a campaign that maximizes the click-through-rate (CTR). Ideally the firm would tailor the advertisement to each consumer based on his on-line footprint and all the accumulated data ($t^n \in T$ in our model). Realistically this is impossible and firms can only generate a limited number of advertisement variants. Let us denote this (exogenously given) bound by $K$. How should a firm partition the set of consumers and what is the optimal variant for each segment? This is the (mechanism) design challenge we are interested in.

We assume that the firm has the ability to identify the optimal advertisement (highest CTR) from the set of all possible variants for any given consumer, $t \in T$. Thus, we can actually identify an ad variant as a point in $T$ as well, where variant $t$ actually refers to the advertisement targeted ideally at consumer $t$.

We will further assume that the further away a consumer is from $t$ the less likely he is to click the ad $t$. To formalize this let $|t - t'|$ denote the $l_1$ distance between any two types of consumers, $t$ and $t'$ (or between a consumer type and an ad variant).[8] The range of values of $|t - t'|$ is therefore $[0, I]$. Let $P : [0, I] \to [0, 1]$ be an arbitrary decreasing function. $P(|t - t'|)$ denotes the CTR, i.e., the probability that a consumer of type $t$ will click on an ad that is designed for type $t'$. We assume $P$ is a Lipschitz function and so there exists some $B > 0$ such that

$$B \cdot |x - x'| \geq |P(x) - P(x')| \text{ for all } x, x' \in [0, I].$$

One way to interpret this assumption is that consumers exhibit some tolerance to product variants which do not match them perfectly. A high value of $B$ corresponds to low tolerance while a low value corresponds to high tolerance.[9]

**The segmentation challenge.** Consumers report their attribute vector and then, in turn, the firm chooses (up to) $K$ points in $T$ so as to maximize the CTR. The consumers, aware of the segmentation scheme, might be strategic about what they report.

Formally, the set of alternatives available to the firm (choosing $K$ variants) is $S = T^K$. For each $s = (s^k)_{k=1}^K \in S$ and for any $t^n \in T$ let $Q(t^n, s) \in argmin_{s^k} |s^k - t^n|$ denote an option closest to $t^n$ in $s$.[10] A vector $t = (t^n)_{n=1}^N \in T^N$ and a choice of $s \in S^K$ of $K$ ad variants induces the vector of probabilities
$$\{P(|t^n - Q(t^n, s)|)\}_{n=1}^N \in [0,1]^N.$$

---

[8]Formally, if $x \in \mathbb{R}^I$ then $|x| = \sum_{i=1}^I |x_i|$. Note that $|x| \in [0, I]$ for $x \in [0,1]^I$.
[9]The model can be generalized so that the probability function $P$ can be different for different consumers. Let us denote by $P^n$ the probability function corresponding to consumer $n$ and by $B^n$ the corresponding Lipschitz coefficient. Our results extend to this setting once we set $B = \max_n B^n$.

The goal of the mechanism is to choose some $s \in S$ in order to maximize the *average CTR* which is defined as
$$F(t,s) = \frac{1}{N} \sum_{n=1}^N P(|t^n - Q(t^n, s)|).$$

A consumer's utility is a function of the product suitability. Consumers view advertisements that are tailored for them as informative while other advertisements can be regarded as spam. Formally, the utility of a consumer of type $t$ when exposed to an advertisement tailored to type $t'$ is $u(t, t') = V(|t - t'|)$, where $V : [0, I] \to [0, 1]$ is an arbitrary decreasing function calibrated so that $V(0) = 1$. We assume that there exists some $c > 0$ such that
$$c \cdot |x - x'| \leq |V(x) - V(x')| \text{ for all } x, x' \in [0, I].$$

The coefficient $c$ captures the 'intolerance to spam' of consumers.[11]

Note that in our model both the consumer's utility and the firm's utility decrease the further away the true type of the consumer is from the assigned advertisement. Apart from this common feature we assume no other common ground between these two.

**Mechanisms.** Let $\Delta = \Delta(S)$ be the set of all probability distributions over $(S, \mathcal{B})$, where $\mathcal{B}$ is standard Borel $\sigma$-field over $S = T^K$, which is just the $I \cdot K$- dimensional unit cube . A (random) mechanism is a function $M : T^N \to \Delta$. Given a vector of consumer types, $t \in T^N$, a vector $b \in T^N$ of announcements and a mechanism $M$ we denote by
$$U_M^n(t^n, b) = E_{s \sim M(b)}(u(t^n, Q(b^n, s))) = E_{s \sim M(b)} V(|t^n - Q(b^n, s)|)$$
the expected utility of consumer $n$ and by
$$F_M(t, b) = E_{s \sim M(b)}(F(t,s)) = E_{\sigma \sim M(b)} \frac{1}{N} \sum_{n=1}^N P(|t^n - Q(t^n, s)|)$$
the expected CTR, where $s$ is randomly drawn according to $M(b) \in \Delta$.

---

[10] In case there is more than one such option we take an arbitrary tie breaking rule such as choosing the one with the smaller index. This tie breaking rule is without loss of generality.

[11] The model can be generalized so that the spam function $V$ can be different for different consumers. Let us denote by $V^n$ the probability function corresponding to consumer $n$ and by $c^n$ the corresponding intolerance to spam coefficient. Our results extend to this setting once we set $c = \min_n c^n$. Our scheme does not rely on designer's familiarity with the specific structure of the functions $V^n$ besides knowing the coefficient $c$ (or at least some reasonable lower bound on it).

**Implementation in undominated strategies.** A (pure) strategy for player $n$ is a function $f^n : T \to T$ that maps $n$'s type into an announced type (announcement). Let $f = (f^n)_{n=1}^N$ denote a vector of strategies. The strategy $f^n$ is *weakly dominated* if there exists some strategy $g^n$ such that for any strategy tuple, $f^{-n}$, of the other players and for any $t \in T^N$, $U_M^n(t^n, f(t)) \leq U_M^n(t^n, (g^n, f^{-n})(t))$, with at least one of the inequalities being strict. If $f^n$ is not weakly dominated we refer to it as *undominated*.

The solution concept is that of undominated strategies which assumes very little in the way of players' rationality, yet is adequate for prior-free environments. We say that $M$ $\eta$-*implements $F$ in undominated strategies* if for any vector of types, $t \in T^N$ and any tuple of undominated strategies, $f$,
$$F_M(t, f(t)) \geq \max_{s \in S} F(t, s) - \eta.$$

## 2.1 Road map of results

The rest of Section 2 is devoted to the construction of the segmentation mechanism. Before we dive into the technicalities let us provide the intuition behind our construction. Our mechanism is a convex combinations of two direct mechanisms, each with its own strengths and weaknesses, such that when properly combined yield the desired result.

The first of the two is the quantal response mechanism. This mechanism has two interesting properties. On the one hand if players are almost truthful then the mechanism outputs a social outcome that is nearly optimal for a large enough population (see Theorems 1,2). On the other hand, although players need not be truthful there is a bound on the gain from misreporting and this bound is linear in the amount of misreporting (Theorem 3).

The second mechanism is an oblivious mechanism which trivially induces truth telling. More so, the loss from misreporting in the mechanism is at least quadratic in the amount misreporting (Lemma 1). The main weakness of the oblivious mechanism is that it can be arbitrarily inefficient.

By properly combining the two mechanisms we get a mechanism where players will not lie by too much about their type (recall the linear upper bound on the gain of the first mechanism vs. the quadratic lower bound on the loss in the second). In addition, for a large enough population the proper combination will put a low enough weight on the oblivious component and hence the efficiency of the mechanism will be almost entirely derived from that of the quantal response mechanism, which we already know to be almost optimal.

## 2.2 A Quantal Response Mechanism

A quantal response mechanism is a mechanism which assigns higher probabilities to better social alternatives, for a given vector of announcements. One particular family of quantal response mechanisms, $\{M_\epsilon : \epsilon > 0\}$, is defined as follows. Let $\lambda$ denote the Lebesgue measure over the unit interval. For any $\bar{S} \subset S$ such that $\bar{S}$ is in the $\sigma$-field $\mathcal{B}$ we set:

$$M_\epsilon(t)(\bar{S}) = \frac{\int_{\bar{S}} e^{n\epsilon(F(t,s))} d\lambda(s)}{\int_S e^{n\epsilon(F(t,s))} d\lambda(s)}. \tag{1}$$

Such a mechanism has previously been introduced by McSherry and Talwar [33] in the context of privacy preserving mechanisms. In their work McSherry and Talwar refer to it as the *exponential mechanism* and demonstrate its properties in the context of mechanism design. We later leverage the results of McSherry and Talwar to show that the mechanism we eventually propose has desirable privacy preserving properties (Section 4). Further connections between the exponential mechanism and mechanism design is also the subject of Nissim, Smorodinsky and Tennenholtz [35], Nissim, Orlandi, and Smorodinsky [34] and Huang and Kannan [25]. We note that the setting of Huang and Kannan differs from ours as their mechanism uses monetary transfers which we avoid.

We now turn to argue that exponential mechanisms have two notable properties. First, deviations from truthfulness can offer very limited gain to a deviating player. Second, small deviations from truthfulness induce an almost optimal outcome. We begin with a formal statement of the latter observation. In fact, whenever players are truthful, the following holds:

**Theorem 1 ([33])** *For any $t \in T^N$, $F_{M_\epsilon}(t,t) \geq \max_s F(t,s) - \frac{3}{N\epsilon} \ln\left(e + (N\epsilon)^{K+1}\right)$.*

Theorem 1 states that if all consumers are truthful then the exponential mechanism with parameter $\epsilon$ would induce a potentially sub optimal outcome. However the expected decrease in social welfare, compared with the first best option, would be of the order of magnitude of $\frac{K \ln(N\epsilon)}{N\epsilon}$, which diminishes to zero as $N$ increases, keeping $\epsilon$ and $K$ fixed. (Later we will set $\epsilon$ to be a function of $N$, but keep $N\epsilon$ increasing with $N$ so as to keep $\frac{K \ln(N\epsilon)}{N\epsilon}$ diminishing overall.)

The limited sub-optimality resulting from truthful reporting can further deteriorate if players are not truthful, however this further deterioration is also bounded. The new bound takes into account the Lipschitz coefficients, $B$ and $c$, previously introduced. In order to save on notation we assume, in what follows, that without loss of generality $c = \frac{1}{B}$ (otherwise set $B = \max\{B, \frac{1}{c}\}$):

**Theorem 2** *Let $\beta \geq 0$. If $\forall n \ |b^n - t^n| \leq \beta$ then $F_{M_\epsilon}(t,b) \geq \max_s F(t,s) - 2B\beta - \frac{3}{N\epsilon} \ln\left(e + (N\epsilon)^{K+1}\right)$.*

The proof of Theorem 2 is deferred to the appendix.

In words, Theorem 2 states that if consumers mis-report their information by (at most) $\beta$ then the bound on the efficiency loss of the exponential mechanism is of the order of magnitude of $B\beta + \frac{K \ln(N\epsilon)}{N\epsilon}$.

The next theorem provides an upper bound on the gain of a consumer by misreporting her true type to the quantal response mechanisms:

**Theorem 3** *If* $\epsilon \leq \frac{B}{2I}$ *then for any* $n$, *any* $b^n, t^n \in T$ *and any* $b^{-n} \in T^{N-1}$,

$$U^n_{M_\epsilon}(t^n, (b^n, b^{-n})) \leq U^n_{M_\epsilon}(t^n, (t^n, b^{-n})) + 4\epsilon B|t^n - b^n|.\text{[12]}$$

The proof of Theorem 3 is deferred to the appendix.

Thus, a consumer may gain by reporting $b^n$ when his true type is $t^n$. However, this gain is bounded by $4\epsilon B|t^n - b^n|$, regardless of what other consumers report.

## 2.3 An Oblivious Mechanism

An *oblivious mechanism* is a mechanism that chooses the $K$ segments while ignoring consumers' announcements. The only use made of consumer announcements is to assign each consumer to a specific segment, once such $K$ segments have been chosen. Therefore if the mechanism assigns each consumer to the segment that is optimal vis-a-vis their announcement then it must be incentive compatible.

The specific oblivious mechanism we have in mind is one which chooses only two points in the space $T$ which value differs only on one coordinate, $i \in I$. For all other $I - 1$ coordinates the values of both products are set to zero. The two distinct values are chosen uniformly in $[0, 1]$, but are restricted to a grid, with step size $x$, over the interval. In fact, we consider a randomized version of this, where the choice of coordinate $i$ and the grid step size $x$ are randomly chosen.

In more detail, we study a family of (random) oblivious incentive compatible mechanisms, parameterized by a positive integer $\bar{m}$. These oblivious mechanisms choose two products which are two random neighbors in some grid on $T = [0,1]^I$. The grid step-size is $I2^X$, where $X$ is chosen at random. Once a grid is given the two products are chosen such that all $I$ attributes are assigned the value zero, except for a single random attribute chosen randomly and uniformly from $\{1, \ldots, I\}$. The value of this remaining attribute for the two neighboring products is chosen uniformly from all corresponding points on the grid. We provide a formal construction below.

First, let $X$ be a random variable which takes on values in $X \in \{1, 2, 3, \ldots, \bar{m}\}$ according to a uniform distribution. Let $\hat{I}$ be another random variable that takes values in $\{1, \ldots, I\}$ according to a uniform distribution and let $\hat{g}$ be a third random variable that takes values in $\{1, 2, \ldots, I2^X - 1\}$ according to a uniform distribution. Now set $\beta = \beta(X, \hat{I}, \hat{g}) = (\vec{0}_{-\hat{I}}, \frac{\hat{g}}{I2^X})$ and $\gamma = \gamma(X, \hat{I}, \hat{g}) = (\vec{0}_{-\hat{I}}, \frac{\hat{g}+1}{I2^X})$. Note that, conditional on $X$, $\beta$ is chosen uniformly from a set of size $I^2 2^X$.

Given a realization of the three random variables $X, \hat{I}, \hat{g}$ the oblivious mechanism chooses the social alternative $s = s(\beta, \gamma) = s(\beta(X, \hat{I}, \hat{g}), \gamma(X, \hat{I}, \hat{g}))$ where it produces the optimal advertisements for types $\beta$ and ($K$-1 copies of) $\gamma$.

---

[12]Recall that $B$ is the Lipschitz coefficient of the click probability function and the consumers' utility function.

We denote this mechanism by $M^{(obl\ \bar{m})}$.

It is easy to verify that an oblivious mechanism is necessarily truthful. However, this specific mechanism yields a lower bound on the utility loss from mis-reporting:

**Lemma 1** *If* $|b^n - t^n| \geq 2^{-(\bar{m}-1)}$ *then* $U^n_{M^{(obl\ \bar{m})}}(t^n, t) \geq U^n_{M^{(obl\ \bar{m})}}(t^n, (b^n, t^{-n})) + \frac{\|t^n - b^n\|_1^2}{\bar{m}8I^2B}$.

The proof of Lemma 1 is deferred to the appendix. To intuit the construction, consider the one dimensional case ($I = 1$). Agent $n$ loses in utility when it is assigned to the advertisement in $\{\beta, \gamma\}$ that is farther from $t^n$. In particular, when $t^n \leq \beta$ and $b^n \geq \gamma$ (or $b^n \leq \beta$ and $t^n \geq \gamma$) the agent loss in utility is $\gamma - \beta$. To maximize the expectation from this kind of utility loss, the mechanism designer wishes to set $\gamma - \beta$ to be of the same magnitude as $|t^n - b^n|$. The designer, however, does not know $|t^n - b^n|$, and furthermore, the magnitude of deviation may differ among agents. Choosing $\gamma - \beta$ according to an exponential scale addresses these two issues simultaneously as for every agent with $|b^n - t^n| \geq 2^{-(\bar{m}-1)}$, the chances of selecting the scale that approximately maximizes her expected utility loss is at least $1/\bar{m}$.

## 3  Main Result - Implementation in undominated strategies

Given a parameter $0 \leq q \leq 1$ let $M_{q,\epsilon,\bar{m}} = (1-q)M_\epsilon + qM^{(obl\ \bar{m})}$ be the random mechanism resulting from flipping a $(q, 1-q)$ biased coin and resorting to the oblivious mechanism, $M^{(obl\ \bar{m})}$, in one case and to the quantal response mechanism, $M_\epsilon$, in the other case.[13]

Note that as $q$ tends to one the mechanism essentially becomes incentive compatible while terribly inefficient. On the other hand for low values of $q$ we may lose incentive compatibility but gain efficiency. In a similar way, small values of $\epsilon$ imply more incentive compatibility while inducing less efficiency. In what follows we pursue some 'golden path' and determine the value of the parameters of the mechanism such that we have both incentive compatibility and (almost) efficiency.

Given a population size $N$, an attribute space of dimension $I$ and the segmentation parameter $K$ we set $\epsilon = \frac{1}{N^{2/3}}$, $\bar{m} = \lceil \log_2\left(\frac{N^{1/3}}{\ln N}\right) \rceil$ (where $\lceil x \rceil$ denotes the smallest integer greater or equal $x$), and $q = 32B^2I^2\epsilon\bar{m}2^{\bar{m}}$. Denote by $\hat{M} = \hat{M}(N, K) = M_{q,\epsilon,\bar{m}}$ the mechanism resulting from this choice of parameters.

Our main result is:

---

[13] An alternative mechanism is to compute the segmentation for both $M_\epsilon$ and $M^{(obl\ \bar{m})}$ resulting in two segments for each consumer ($K + 2$ segments overall). Then, for each consumer separately and independently, flip a $(q, 1-q)$ coin in order to determine which of the two segments to apply. The advantage is that the variance in the objective is reduced, as we never end up with an arbitrary segment for all consumers at once. The disadvantage is that this mechanism is somewhat wasteful in the number of segments, which is $K + 2$ instead of $K$. Our main result holds for this alternative mechanism as well.

**Theorem 4** *For any $I, K$ there exists $\kappa = \kappa(I, K)$ and $N_0$ such that $\hat{M}$ $\kappa \frac{\ln N}{N^{1/3}}$-implements $F$ in undominated strategies for all $N > N_0$.*

In particular $\kappa = (32B^2 I^2 + 2 + 3(K + 2))$, where $B$ is the Lipschitz coefficient of the function $P$.

**Proof of Theorem 4:**

Observe that for any $I$ and $K$ there exists $N_q = N_q(I, K)$ such that $q < 1$ for all $N > N_q$ and hence the mechanism is well defined.

Now assume $i$ reports $b_i$ such that $|b^i - t^i| \geq 2^{-(\bar{m}-1)}$. Therefore, given the choice of parameters $q \frac{|t^i - b^i|}{\bar{m} I^2 2^{2I+3}} \geq 4\epsilon B$. Applying Lemma 1 and Theorem 3 we deduce that:

$$U^n_{\hat{M}}(t^i, t) - U^n_{\hat{M}}(t^i, (b^i, t^{-i})) =$$

$$(1-q)\left[U^n_{M_\epsilon}(t^i, t) - U^n_{M_\epsilon}(t^i, (b^i, t^{-i}))\right] + q\left[U^n_{M^{(obl\ \bar{m})}}(t^i, t) - U^n_{M^{(obl\ \bar{m})}}(t^i, (b^i, t^{-i}))\right] \geq$$

$$-(1-q)4\epsilon B|t^i - b^i| + q\frac{\|t^n - b^n\|_1^2}{\bar{m} 8 I^2 B} \geq q\frac{\|t^n - b^n\|_1^2}{\bar{m} 8 I^2 B} - 4\epsilon B|t^i - b^i|$$

Replacing with $q = 32 B^2 I^2 \epsilon \bar{m} 2^{\bar{m}}$:

$$U^n_{\hat{M}}(t^i, t) - U^n_{\hat{M}}(t^i, (b^i, t^{-i})) \geq \frac{32 B^2 I^2 \epsilon \bar{m} 2^{\bar{m}} |t^i - b^i|^2}{\bar{m} 8 I^2 B} - 4\epsilon B|t^i - b^i| =$$

$$= 4 B \epsilon 2^{\bar{m}} |t^i - b^i|^2 - 4\epsilon B |t^i - b^i| = 4\epsilon B |t^i - b^i| \left(2^{\bar{m}} |t^i - b^i| - 1\right) > 0.$$

In words, truthfulness dominates any strategy where $|b^i - t^i| \geq 2^{-(\bar{m}-1)}$. Therefore, we analyze the outcome of the mechanism when all consumers comply with the requirement $|b^i - t^i| < 2^{-(\bar{m}-1)}$

There exists some integer $N_\alpha = N_\alpha(I, K)$ such that $\frac{2}{N\epsilon} \ln\left(e + (N\epsilon)^{K+1}\right) \leq 0.5$ for all $N > N_\alpha$ as required in the proof of Theorem 1. In addition, there exists $N_{\bar{m}} = N_{\bar{m}}(I, K)$ such that $\bar{m} \leq \ln N$ for all $N > N_{\bar{m}}$. Hereinafter we assume that $N > \max(N_q, N_\alpha, N_{\bar{m}})$.

There are two sources for the sub-optimality of $\hat{M}$:

1. As the CTR is bounded between 0 and 1, the introduction of an oblivious mechanism with probability $q$ introduces an expected additive error of at most $q = 32 B^2 I^2 \epsilon \bar{m} 2^{\bar{m}}$. Noting that $2^{\bar{m}} \leq \frac{N^{1/3}}{\ln N}$ and substituting for $\epsilon$ we get that $q \leq 32 B^2 I^2 \frac{1}{N^{2/3}} \bar{m} \frac{N^{1/3}}{\ln N}$. Recall that $\bar{m} \leq \ln N$ and so $q \leq \frac{32 B^2 I^2}{N^{1/3}} \leq \frac{32 B^2 I^2 \ln N}{N^{1/3}}$

2. By Theorem 2 the quantal response mechanism introduces an additive error of

$$\frac{B}{2^{(\bar{m}-1)}} + \frac{3}{N\epsilon} \ln\left(e + (N\epsilon)^{K+1}\right).$$

We consider each of the two additive errors:

(a) As for the first factor, note that $2^{\bar{m}-1} = \frac{1}{2} \cdot 2^{\bar{m}} \geq \frac{1}{2} \cdot \left(\frac{N^{1/3}}{\ln N}\right)$ which implies that this factor error is bounded by $\frac{2\ln N}{N^{1/3}}$.

(b) Note that the second factor, $\frac{3}{N\epsilon} \ln\left(e + (N\epsilon)^{K+1}\right) \leq \frac{3(K+2)}{N\epsilon} \ln(N\epsilon)$ and substituting for $\epsilon$, we get that there exists $N_1 = N_1(K)$ such that for all $N > N_1$ this additive error is bounded by $\frac{3(K+2)}{N^{1/3}} \ln N$.

Setting $N_0 = \max(N_q, N_\alpha, N_{\bar{m}}, N_1)$, we get that for all $N > N_0$ the total additive error is bounded by

$$\left(32B^2I^2 + 2 + 3(K+2)\right) \frac{\ln N}{N^{1/3}}$$

**QED**

One criticism of the proposed mechanism is that it hinges on the exponential mechanism, which is computationally demanding.[14]

Note, however, that any mechanism that satisfies Theorems 1,2,3 could successfully replace the exponential mechanism. Theretofore, if future work will lead to the discovery of such a mechanism then it could easily replace the exponential mechanism (perhaps with an appropriate change in the constants that refer to the accuracy and the population size). In that sense our main result can be reinterpreted as a black-box reduction that can be applied to any differentially private mechanism for the segmentation problem to get the desirable incentive properties.

# 4 Privacy Issues

Privacy has emerged in the recent years as a public concern in markets where personal information is gathered and used for commercial goals in general and segmentation in particular. The concerns around privacy refer to the way personal data is saved, who has access to it, how it can be used in future interactions and whether or not it is forwarded to additional entities. Most of these concerns are either handled using legal tools or cryptographic tools. However, there is one aspect of privacy which can be treated neither with legal nor with cryptographic tools. As the outcome of the mechanism (in particular the segments in our setting) is made public, one could argue that this information could be used to derive some of the private input provided by the consumers and hence jeopardize privacy. This concern can only be addressed through the design of the mechanism. That is, the mechanism should be constructed in a way that makes it difficult to reverse engineer and individual data from the public outcome.

---

[14] For some settings of the objective function $F$, sampling from the exponential mechanism amounts to breaking cryptographic assumptions. An example is when the exponential mechanism is used in an implementation of (differentially private) data release [17, 43, 44]. It is not known whether this applies also to the current setting of $F$ and so the question of a computationally efficient implementation of our mechanism is open. Attempts to sample from the exponential mechanism include the use of Markov Chain Monte Carlo (MCMC) sampling as in [11]. A subtle issue with this approach is that while the MCMC converges to the distribution of the exponential mechanism and hence satisfies differential privacy 'in the limit', it is only approximate and potentially not satisfying differential privacy if the MCMC is run for only a finite number of steps.

To be more specific the concern is that even if consumers' personal data is kept confidential and the communication channels between the consumers and the mechanism are encrypted, one may still learn something about individuals just from observing the outcome of the mechanism. In particular, in our setting, by observing the choice of $K$ segments one may deduce something about the private information of some consumers. Thus, privacy should be carefully thought of when designing mechanisms.

How should one account for the privacy loss of a mechanism? In particular, for any given customer what is the privacy loss she faces? Conceptually, to ensure privacy, one should design the mechanism so that no individual input has an observable impact on the mechanism's outcome. In other words, if one compares the two outcomes of the mechanism: that which is derived from all the data, the specific customer's data included, and that which is derived from the same data, absent the specific customer's data, then the gap should be minute. In this section we focus on a particular methodology to account for privacy and the aforementioned gap. For this specific methodology we argue that the mechanism provided complies with privacy constraints.

**Differential privacy**: The specific methodology we adopt for accounting for privacy is one that has been developed over the course of the last decade in the theoretical literature of computer science and is known as *differential privacy*. We adopt this methodology as it the current academic best practice used for defining privacy. Differential privacy, due to Dwork, McSherry, Nissim, and Smith [16]. is formalized as follows:

Let $M : T^N \to \Delta(S)$ be a mechanism. For $t \in T^N$ and a measurable set $\hat{S} \subset S$ we denote by $M(t)(\hat{S})$ the probability that $M(t)$ chooses some vector of $K$ segments in $\hat{S}$. A pair of type vectors, $t, \hat{t}$ in $T^N$, are called *neighbors* if they differ on a single coordinate. Formally, $|\{i : t_i = \hat{t}_i\}| = N - 1$.

**Definition 1 ([16])** *M provides $\epsilon$-differential privacy if for any measurable $\hat{S} \subset S$, any pair of neighbors $t, \hat{t} \in T^N$, $M(t)(\hat{S}) \leq e^\epsilon \cdot M(\hat{t})(\hat{S})$.*

This inspires the following:

**Definition 2** *The privacy score of a mechanism $M : T^N \to \Delta(S)$ is $\rho(M) = \max \ln \frac{M(t)(\hat{S})}{M(\hat{t})(\hat{S})}$, where the maximum is taken over all measurable set $\hat{S} \subset S$ and all neighbors $t, \hat{t} \in T^N$.*

We draw the reader's attention to the following: (1) The higher the score the less privacy is preserved. (2) The above definition works in our non-Bayesian setting, where no prior over the vector of types is specified. (3) A worst-case spirit underlies this definition. Thus, if we can guarantee a low privacy score with such a worst-case definition then we can do so for alternative scores. Indeed, the following theorem provides an upper bound on the privacy score that diminishes to zero with the size of the population.

Let us consider the privacy score of the mechanism we propose in our main Theorem:

**Theorem 5** *The privacy score of the mechanism $\hat{M}$ satisfies $\rho(\hat{M}) \leq \frac{2}{N^{2/3}}$.*

The proof of Theorem 5 hinges on the following observation (which proof is straightforward and is omitted):

**Lemma 2** *Assume mechanism $M_i$ has a privacy score of $\epsilon_i$ and let $M = \alpha M_1 + (1-\alpha)M_2$, where $0 \leq \alpha \leq 1$. Then $\rho(M) \leq \max_i \epsilon_i$.*

**Proof of Theorem 5:** Recall that $\hat{M}$ combines an exponential mechanism, $M_\epsilon$, and an oblivious mechanism. McSherry and Talwar [33] show that $\rho(M_\epsilon) \leq 2\epsilon$. Recall that $\epsilon = \frac{1}{n^{2/3}}$ in our case and so $\rho(M_\epsilon) \leq \frac{2}{n^{2/3}}$. The oblivious mechanism ignores players' inputs and so has a privacy score of zero. Now invoke Lemma 2 to finalize the proof.

**QED**

At this point a disclaimer is clearly due. The actual segmentation mechanism does more then produce a desired set of segments. It also allocates each customer to one such segment. If such allocation is publicly announced then our analysis on privacy is clearly flawed as the associated segment provides valuable information on the customer's input. Therefore the above analysis hinges on the assumption that each customer is *discreetly* associated with a segment. Clearly, without this assumption privacy cannot be guaranteed.[15]

---

[15]The notion of *Joint differential privacy*, see [28], captures privacy considerations in such cases.

# 5   Can the Mechanism be Simplified?

One desirable property of mechanisms is simplicity. A simple mechanism implies that participants have an easy time understanding the 'rules of the game'. We do not pursue a formal notion of simplicity however we admit that the proposed mechanism is not simple but is rather complex. The mechanism is composed in a very specific way from two more primitive mechanisms, of which one - the quantal response mechanism - is quite complex in its own right. A valid concern about complex mechanisms in general and ours in particular is whether they may deter participation. Thus, it would have been nice to achieve similar efficiency and privacy results in a simpler mechanism.

One could argue that as we focus on large populations, individuals are likely to be non influential and so eliciting the true information should be straightforward and is possible without the complexity we introduce. It is beyond our knowledge to argue that this intuition is generally incorrect, however we demonstrate it is a problematic argument via an analysis of the following two natural simplifications of our mechanism:

- **Variant 1** - Our proposed mechanism randomizes between a quantal response mechanism and an oblivious one, where the probability of the latter, $q = q(N)$, diminishes as the population size, N, grows. In variant 1 we maintain such a randomization but replace the quantal response mechanism with a mechanism that produces the naive first best outcome (assigns probability one to the optimal segmentation).

- **Variant 2** - A variant which only uses the quantal response mechanism and ignores the oblivious mechanism.

Unfortunately, neither of these two variants has the desired properties as we demonstrate below.

**A counterexample for Variant 1**: In the next example we provide an equilibrium for the 'segmentation game' with complete information derived from variant 1 of our mechanism. In this equilibrium a fixed proportion of the consumers are not truthful and consequently the segmentation is sub-optimal. Furthermore the gap from optimality does not diminish as the population size grows.

**Example 1** *Let $K = 2$ and assume the consumer's satisfaction and the CTR diminish at a rate of $V(x) = P(x) = x^2$, where $x$ is the deviation from truthfulness. Let the set of possible types be single dimensional ($I = 1$) and consider the scenario with $12N$ consumers, of which $N$ are of type $t = 0$, $N$ of type $t = 0.6$ and $10N$ of type $t = 1$. We assume, without loss of generality, that if there is more than a single optimal segmentation given players' announcements, then the tie breaking rule will choose for minimize the lower end product.*

*Consider the following vector of announcements: Consumers with extreme types ($t = 0, 1$) announce truthfully whereas consumers with type $t = 0.6$ announce $y$. It turns out that if $y$ solves the equation $(\frac{y}{1-y})^2 = \frac{20}{11}$ then the following two segmentations are optimal vis-a-vis the announcements:*

1. Set $s^1 = \frac{y}{2}, s^2 = 1$ and lump the $2N$ consumers with types $t = 0, 0.6$ around the first product.

2. Set $s^1 = 0, s^2 = \frac{y+10}{11}$ and lump the $11N$ consumers with types $t = 0.6, 1$ around the second product.

The social welfare (CTR) from the first option is $\frac{N[1-(\frac{y}{2})^2]+N[1-(\frac{y}{2})^2]+10N}{12N}$ whereas the social welfare from the second option is $\frac{N+N[1-(y-(\frac{y+10}{11}))^2]+10N[1-(1-\frac{y+10}{11})^2]}{12N}$. Equating these two results in the equation $(\frac{y}{1-y})^2 = \frac{20}{11}$ which is how we chose $y$ to begin with.

Given our tie breaking rule the segmentation chosen will be the former

Note that the only potentially profitable deviation by a single player is for a player of type $0.6$ to announce a value greater than $y$. However, any such deviation will result in a segmentation that is very similar to the latter option. For the true type ($t = 0.6$) this results in a discontinuous drop in utility. This discontinuity implies that the vector of announcements forms a Nash equilibrium for variant 1 when the oblivious mechanism is chosen with a sufficiently small probability.

To finish the example we compute the CTR in equilibrium which is $\frac{N(1-(\frac{y}{2})^2)+N[1-(0.6-(\frac{y}{2})^2)]+10N}{12N} \approx \frac{12-0.18}{12}$ and compare it with the optimal CTR which is $\frac{N+N[1-(\frac{10+0.6}{11}-0.6)^2]+10N[1-(1-\frac{10+0.6}{11})^2]}{12N} \approx \frac{12-0.1455}{12}$. This inefficiency gap does not depend on $N$ and in particular does not diminish as the population size grows.

**A counterexample for Variant 2**: In the next example we provide an equilibrium for the 'segmentation game' with complete information derived from variant 2 of our mechanism. Recall that the second variant involves dropping the oblivious mechanism and using only the exponential component. In the following example the equilibrium involves half the consumers not being truthful. The resulting segmentation is consequently sub-optimal and the gap from optimality does not diminish as the population size grows.

**Example 2** Let $K = 1$ and assume the consumer's satisfaction and the CTR diminish at a rate of $V(x) = P(x) = x^2$, where $x$ is the deviation from truthfulness. Let the set of possible types be single dimensional ($I = 1$) and consider the scenario with $2N$ consumers, of which $N$ are of type $t = 0$ and $N$ of type $t = 0.5$. The equilibrium (for sufficiently large $N$) is for type $0$ consumers to be truthful and for type $0.5$ consumers to announce $b = 1$. We leave it to the reader to verify this is indeed an equilibrium and that the efficiency loss from optimality is roughly $[1-2\frac{1}{4}^2]-[1-\frac{1}{2}^2] = \frac{1}{8}$.

# 6 Summary

The idea of considering an optimal set of $K$ representative agents may be seen as a contribution to the development of economic methodology. The use of representative agents, where all customers are considered alike a "typical" one, has been central to the economic literature [21]. In contrast, game theoretic models, which model each and every agent separately may be too challenging. What we implicitly consider here is a middle ground where a large economy can be modeled vis-a-vis studying $K > 1$ representative agents. Instead of ending up with a single economic policy based on the single representative agent model as in the standard approach, models that deal with $K > 1$ representative agents may propose a menu of policies (e.g., tax policies) coupled with a means for assigning each individual to one of the policies.

# 7 Appendix - Proofs

## 7.1 Proof of Theorem 1

We restate Theorem 1 for convenience:

**Theorem 1**: *For any $t \in T^n$, $F_{M_\epsilon}(t,t) \geq \max_s F(t,s) - \frac{3}{n\epsilon} \ln \left(e + (n\epsilon)^{K+1}\right)$.*

Before proving the theorem we cite a technical observation due to McSherry and Talwar [33]. To do so we introduce the following notation. For $0 \leq \alpha \leq 1$ let $S_\alpha = S_\alpha(t) = \{s \in S : F(t,s) \geq \max_s F(t,s) - \alpha\}$, and $\bar{S}_\alpha = \bar{S}_\alpha(t) = S \setminus S_\alpha$.

Let $\lambda$ denote the Lebesgue measure over $[0,1]^{IK}$.

**Lemma 3 (McSherry and Talwar [33])** *If $\alpha \geq \frac{1}{n\epsilon} \ln \left(\frac{\max_s F(t,s)}{\alpha \lambda(S_\alpha)}\right)$ then $F_{M_\epsilon}(t,t) \geq \max_s F(t,s) - 3\alpha$.*

We include the proof for completeness.

**Proof**: Note first that

$$\begin{aligned}
M_\epsilon(t)(\bar{S}_{2\alpha}) &\leq \frac{M_\epsilon(t)(\bar{S}_{2\alpha})}{M_\epsilon(t)(S_\alpha)} = \frac{\int_{\bar{S}_{2\alpha}} e^{n\epsilon(F(t,\bar{s}))} d\bar{s}}{\int_{S_\alpha} e^{n\epsilon(F(t,\bar{s}))} d\bar{s}} \leq \\
&\leq \frac{\int_{\bar{S}_{2\alpha}} e^{n\epsilon(\max_s F(t,s) - 2\alpha)} d\bar{s}}{\int_{S_\alpha} e^{n\epsilon(\max_s F(t,s) - \alpha)} d\bar{s}} = e^{-n\epsilon\alpha} \cdot \frac{\lambda(\bar{S}_{2\alpha})}{\lambda(S_\alpha)} \leq \frac{e^{-n\epsilon\alpha}}{\lambda(S_\alpha)},
\end{aligned}$$

where the first inequality follows from $M_\epsilon(t)(S_\alpha) \leq 1$, the second inequality follows from the definition of $\bar{S}_{2\alpha}$ and $S_\alpha$, and the third inequality follows from $\lambda(\bar{S}_{2\alpha}) \leq 1$.

Recall that $\alpha \geq \frac{1}{n\epsilon} \ln\left(\frac{\max_s F(t,s)}{\alpha \lambda(S_\alpha)}\right)$, therefore implying $\frac{e^{-n\epsilon\alpha}}{\lambda(S_\alpha)} \leq \frac{\alpha}{\max_s F(t,s)}$. Combined with the previous inequality proves that $M_\epsilon(t)(\bar{S}_{2\alpha}) \leq \frac{\alpha}{\max_s F(t,s)}$. This, in turn, proves that $M_\epsilon(t)(S_{2\alpha}) \geq 1 - \frac{\alpha}{\max_s F(t,s)}$. In words, $M_\epsilon(t)$ returns $s \in S_{2\alpha}$ with probability at least $1 - \frac{\alpha}{\max_s F(t,s)}$.

Hence,
$$F_{M_\epsilon}(t,t) \geq (\max_s F(t,s) - 2\alpha) M_\epsilon(S_{2\alpha}) \geq$$
$$\geq (\max_s F(t,s) - 2\alpha)(1 - \frac{\alpha}{\max_s F(t,s)}) \geq \max_s F(t,s) - 3\alpha,$$

implying the desired result.

**QED**

Fix $t \in T^n$ and let $\bar{s} \in argmax_s F(t,s)$. Let $\hat{S} = \{s \in S : |s^k - \bar{s}^k| \leq \frac{\alpha}{B}\}$, where $B$ is the Lipschitz coefficient of the function $P$.

**Lemma 4** $\hat{S} \subset S_\alpha$.

**Proof:** Let $\hat{s}$ be an arbitrary element of $\hat{S}$ and let $i$ be an arbitrary buyer. Assume that $Q(t^i, \bar{s})$ is $\bar{s}_k$. Then $P(|t_i - Q(t_i, \hat{s})|) \geq P(|t_i - \hat{s}_k|)$. In addition $P(|t_i - \bar{s}_k|) - P(|t_i - \hat{s}_k|) \leq B(|t_i - \bar{s}_k| - |t_i - \hat{s}_k|) \leq B|\bar{s}_k - \hat{s}_k| \leq \alpha$ (for the first inequality recall that $B$ is the Lipschitz coefficient of $P$ while the second inequality follows from the triangle inequality). Therefore, $P(|t_i - Q(t_i, \hat{s})|) \geq P(|t_i - \bar{s}_k|) - \alpha$.

Summing over all consumers, $F(t, \hat{s}) = \frac{1}{n} \sum_{i=1}^n P(|t_i - Q(t_i, \hat{s})|) \geq \frac{1}{n} \sum_{i=1}^n [P(|t_i - \bar{s}_k|) - \alpha] = F(t, \bar{s}) - \alpha$, which implies $\hat{s} \in S_\alpha$.

**QED**

Using Lemmas 3 and 4 we can now prove Theorem 1:

**Proof of Theorem 1:** Fix $t \in T^n$, let $\bar{s} \in argmax_s F(t,s)$ and set $\hat{S} = \{s \in S : |s^k - \bar{s}^k| \leq \frac{\alpha}{B}\}$. By Lemma 4, $\hat{S} \subset S_\alpha$ and so $\lambda(S_\alpha) \geq \lambda(\hat{S}) \geq (\frac{2\alpha}{B})^K$.

Set $\alpha = \frac{1}{n\epsilon} \ln\left(e + (n\epsilon)^{K+1}\right)$ and let $n_B$ be large enough such that for any $n > n_B$, $2\ln(e + (n\epsilon)^{K+1}) > B$. Therefore

$$\frac{2^K}{(n\epsilon)^{K+1}}[\ln(e + (n\epsilon)^{K+1})]^{K+1} (e + (n\epsilon)^{K+1}) \geq B^K \implies 2^K \alpha^{K+1}(e + (n\epsilon)^{K+1}) \geq B^K \implies$$

$$e + (n\epsilon)^{K+1} \geq \frac{B^K}{\alpha(2\alpha)^K} \implies e + (n\epsilon)^{K+1} \geq \frac{\max_s F(t,s)}{\alpha \lambda(S_\alpha)} \implies$$

$$\frac{\ln(e+(n\epsilon)^{K+1})}{n\epsilon} \geq \frac{\ln(\frac{\max_s F(t,s)}{\alpha\lambda(S_\alpha)})}{n\epsilon} \implies \alpha \geq \frac{1}{n\epsilon}\ln(\frac{\max_s F(t,s)}{\alpha\lambda(S_\alpha)}).$$

Therefore we can apply Lemma 3 and deduce that

$$F_{M_\epsilon}(t,t) \geq \max_s F(t,s) - 3\alpha = \max_s F(t,s) - \frac{3}{n\epsilon}\ln\left(e+(n\epsilon)^{K+1}\right),$$

as required.

**QED**

## 7.2 Proof of Theorem 2

**Theorem 2 restated**: *If $\forall i\ |b^i - t^i| \leq \beta$ then $F_{M_\epsilon}(t,b) \geq \max_s F(t,s) - 2B\beta - \frac{3}{n\epsilon}\ln\left(e+(n\epsilon)^{K+1}\right)$*

**Lemma 5** *For any $s \in S$ and $b, t, \in [0,1]^n$,*

- $|F((b^i, t^{-i}), s) - F((t^i, t^{-i}), s)| \leq \frac{1}{n}B|t^i - b^i|$, and
- $|F(b,s) - F(t,s)| \leq max_i B|t^i - b^i|$.

**Proof:**
$$|b^i - Q(b^i, s)| \leq |b^i - Q(t^i, s)| \leq |b^i - t^i| + |t^i - Q(t^i, s)|.$$

Thus, implying $|b^i - Q(b^i, s)| - |t^i - Q(t^i, s)| \leq |b^i - t^i|$. By a symmetric argument $|t^i - Q(t^i,s)| - |b^i - Q(b^i, s)| \leq |b^i - t^i|$. Therefore:

$$|F(b^i, t^{-i}, s) - F(t^i, t^{-i}, s)| = \frac{1}{n}P(|t^i - Q(t^i, s)|) - P(|b^i - Q(b^i, s)|) \leq$$

$$\leq \frac{1}{n}B||t^i - Q(t^i, s)| - |b^i - Q(b^i, s)|| \leq \frac{1}{n}B|t^i - b^i|,$$

proving the first part of the Lemma.

To prove the second part we introduce the notation $h^i = (b^1, \ldots, b^i, t^{i+1}, \ldots t^n)$, $i = 1, \ldots, n$. From part one we know that $|F(h^{i-1}) - F(h^i)| \leq \frac{1}{n}B|b^i - t^i|$, and so:

$$|F(b,s) - F(t,s)| \leq \sum_{i=1}^n |F(h^{i-1}) - F(h^i)| \leq \sum_{i=1}^n \frac{1}{n}B|b^i - t^i| \leq max_i B|b^i - t^i|.$$

**QED**

**Lemma 6** *If $|b^i - t^i| \leq \beta$ for all $i$ then $|\max_s F(t,s) - \max_s F(b,s)| \leq B\beta$.*

**Proof:** Let $s_t \in S$ be the optimal segmentation for the vector $t$ (formally, $F(t, s_t) \geq F(t, s) \ \forall s \in S$) and similarly let $s_b \in S$ be the optimal segmentation for the vector $b$. ($F(b, s_b) \geq F(b, s) \ \forall s \in S$). The inequality $|t^i - Q(b^i, s_b)| \geq |t^i - Q(t^i, s_b)|$ follows from the definition of the function $Q$. Invoking the triangle inequality:

$$|t^i - b^i| + |b^i - Q(b^i, s_b)| \geq |t^i - Q(b^i, s_b)| \geq |t^i - Q(t^i, s_b)|.$$

In the same vein $|b^i - Q(t^i, s_b)| \geq |b^i - Q(b^i, s_b)|$ and so

$$|t^i - b^i| + |t^i - Q(t^i, s_b)| \geq |b^i - Q(t^i, s_b)| \geq |b^i - Q(b^i, s_b)|.$$

Combining these 2 inequalities:

$$||t^i - Q(t^i, s_b)| - |b^i - Q(b^i, s_b)|| \leq |t^i - b^i|.$$

Therefore,

$$F(b, s_b) - F(t, s_t) \leq F(b, s_b) - F(t, s_b) \leq \frac{1}{n}\sum_{i=1}^{n} |P(|b^i - Q(b^i, s_b)|) - P(|t^i - Q(t^i, s_b)|)| \leq$$

$$\leq \frac{1}{n}\sum_{i=1}^{n} B ||b^i - Q(b^i, s_b)| - |t^i - Q(t^i, s_b)|| \leq \frac{1}{n}\sum_{i=1}^{n} B|t_i - b_i| \leq B\beta.$$

Using symmetrical arguments we conclude that $F(t, s_t) - F(b, s_b) \leq B\beta$. Combining the last inequalities implies that

$$|\max_s F(t, s) - \max_s F(b, s)| = |F(t, s_t) - F(b, s_b)| \leq B\beta,$$

as claimed.

**QED**

**Proof of Theorem 2:** From the second part of Lemma 5 we can conclude, by taking expectation on both sides with respect to the mechanism $M_\epsilon$ at the announcement vector $b$, that $|F_{M_\epsilon}(b,b) - F_{M_\epsilon}(t,b)| \leq B\beta$ and so $F_{M_\epsilon}(t,b) \geq F_{M_\epsilon}(b,b) - B\beta \geq \max_s F(b,s) - \frac{3}{n\epsilon}\ln\left(e + (n\epsilon)^{K+1}\right) - B\beta$, where the last inequality follows from Theorem 1. Invoking Lemma 6 implies that $F_{M_\epsilon}(t,b) \geq \max_s F(t,s) - B\beta - \frac{3}{n\epsilon}\ln\left(e + (n\epsilon)^{K+1}\right) - B\beta = \max_s F(t,s) - 2B\beta - \frac{3}{n\epsilon}\ln\left(e + (n\epsilon)^{K+1}\right)$, as required.

**QED**

## 7.3 Proof of Theorem 3

**Theorem 3 restated**: *If $\epsilon \leq \frac{B}{2I}$ then for any $i$, any $b^i, t^i \in T$ and any $t^{-i} \in T^{n-1}$,*

$$U^n_{M_\epsilon}(t^i, (b^i, t^{-i})) \leq U^n_{M_\epsilon}(t^i, t) + 4\epsilon B |t^i - b^i|.$$

**Proof of Theorem 3:**

$$\begin{aligned}
U^n_{M_\epsilon}(t^i, (b^i, t^{-i})) &= \int_{s \in S} \frac{I - |t^i - Q(b^i, s)|}{I} \, \mathrm{d} M_\epsilon(b^i, t^{-i})(s) \\
&= \int_{s \in S} \frac{I - |t^i - Q(b^i, s)|}{I} \frac{e^{n\epsilon F((b^i, t^{-i}), s)}}{\int_{s' \in S} e^{n\epsilon F((b^i, t^{-i}), s')} \, \mathrm{d} s'} \, \mathrm{d} s \\
&\leq \int_{s \in S} \frac{I - |t^i - Q(b^i, s)|}{I} \frac{e^{n\epsilon \left( F((t^i, t^{-i}), s) + B \frac{|t^i - b^i|}{n} \right)}}{\int_{s' \in S} e^{n\epsilon \left( F(t^i, t^{-i}, s') - B \frac{|t^i - b^i|}{n} \right)} \, \mathrm{d} s'} \, \mathrm{d} s \\
&= e^{2\epsilon B |t^i - b^i|} \int_{s \in S} \frac{I - |t^i - Q(b^i, s)|}{I} \, \mathrm{d} M_\epsilon(t^i, t^{-i})(s) \\
&\leq e^{2\epsilon B |t^i - b^i|} \int_{s \in S} \frac{I - |t^i - Q(t^i, s)|}{I} \, \mathrm{d} M_\epsilon(t^i, t^{-i})(s) \\
&= e^{2\epsilon B |t^i - b^i|} U^n_{M_\epsilon}(t^i, (t^i, t^{-i})).
\end{aligned}$$

It is well known that for any $x \in [0, 1]$, $e^x \leq 1 + 2x$. As $2\epsilon B |t^i - b^i| < 1$ for any $\epsilon < \frac{B}{2I}$ it must be that $e^{2\epsilon B |t^i - b^i|} \leq 1 + 4\epsilon B |t^i - b^i|$. Therefore:

$$U^n_{M_\epsilon}(t^n, (b^n, t^{-n})) \leq (1 + 4\epsilon B |t^i - b^i|) U^n_{M_\epsilon}(t^i, (t^i, t^{-i})) \leq U^n_{M_\epsilon}(t^i, (t^i, t^{-i})) + 4\epsilon B |t^i - b^i|).$$

**QED**

## 7.4 Proof of Lemma 1

**Lemma 1 restated**: If $|b^n - t^n| \geq 2^{-(\bar{m}-1)}$ then $U^n_{M^{(obl\ \bar{m})}}(t^n, t) \geq U^n_{M^{(obl\ \bar{m})}}(t^n, (b^n, t^{-n})) + \frac{\|t^n - b^n\|_1^2}{\bar{m} 8 I^2 B}$.

**Proof of Lemma 1:** For any realization, $i$, of $\hat{I}$ let $\delta_i$ denote

$$\delta_i = \begin{cases} 0 & \text{if } |t^n_i - b^n_i| < 2^{-(\bar{m}-1)}/I \\ |t^n_i - b^n_i| & \text{otherwise} \end{cases}$$

Let $i$ be such that $\delta_i > 0$. Then there exists a realization, $x_i \in \{1, \ldots, \bar{m}-1\}$ of $X$ such that $\frac{1}{I2^{x_i}} < \frac{|t_i^n - b_i^n|}{2} \leq \frac{2}{I2^{x_i}}$.

Assume without loss of generality that $t_i^n < b_i^n$. Conditioning on $\hat{I} = i$ and $X = x_i$ there exists a realization of $\hat{g}$ such that resulting pair of variants, $\beta$ and $\gamma$ satisfy $Q(t^n, s(\hat{X}, \beta, \gamma)) = \beta$, $Q(b^n, s(\hat{X}, \beta, \gamma)) = \gamma$ and $\|\beta - \gamma\|_1 = \frac{1}{I2^{\hat{X}}}$.[16]

For this realization
$$\begin{aligned} u^n(t^n, Q(t^n, s)) - u^n(t^n, Q(b^n, s)) &= V^n(\|t^n - \beta\|_1) - V^n(\|t^n - \gamma\|_1) \geq \\ \frac{1}{B}\|\beta - \gamma\|_1 \geq \frac{1}{B}\frac{1}{I2^{x_i}} &\geq \frac{1}{B}\frac{|t_i^n - b_i^n|}{2} = \frac{1}{B}\frac{\delta_i}{2}. \end{aligned}$$

Thus, conditional on $\hat{I} = i$ and $X = x_i$ the expected decrease in utility is:
$$u^n(t^n, Q(t^n, s)) - u^n(t^n, Q(b^n, s)) = \frac{1}{I2^{x_i}}\frac{1}{B}\frac{\delta_i}{2} \geq \frac{1}{B}\frac{\delta_i^2}{4}.$$

The probability of the realization $X = x_i$ is $\frac{1}{\bar{m}}$ and hence conditional on $\hat{I} = i$:
$$u^n(t^n, Q(t^n, s)) - u^n(t^n, Q(b^n, s)) \geq \frac{1}{\bar{m}B}\frac{\delta_i^2}{4}.$$

On the other hand if $\delta_i = 0$ then clearly a similar inequality holds.

Taking expectation with respect to $\hat{I}$:

$$\begin{aligned} U^n_{M^{(obl\ \bar{m})}}(t^n, t) - U^n_{M^{(obl\ \bar{m})}}(t^n, (b^n, t^{-n})) &\geq \frac{\sum_i \delta_i^2}{\bar{m}4IB} = \frac{\|\delta\|_2^2}{\bar{m}4IB} = \\ \frac{1}{\bar{m}4IB}(\|t^n - b^n\|_2^2 - \sum_{\{i:|t_i^n - b_i^n|<2^{-(\bar{m}-1)}/I\}}(|t_i^n - b_i^n|)^2) &\geq \frac{1}{\bar{m}4IB}(\|t^n - b^n\|_2^2 - I \cdot (2^{-(\bar{m}-1)}/I)^2) \\ &\geq \frac{1}{\bar{m}4IB}\left(\left(\frac{\|t^n - b^n\|_1}{\sqrt{I}}\right)^2 - \frac{2^{-(\bar{2}m-2)}}{I}\right) \\ &= \frac{\|t^n - b^n\|_1^2 - 2^{-(\bar{2}m-2)}}{\bar{m}4I^2B} \geq \frac{\|t^n - b^n\|_1^2}{\bar{m}8I^2B}. \end{aligned}$$

**QED**

# References

[1] D. Abreu and H. Matsushima. "Virtual Implementation in Iteratively Undominated Strategies: InComplete Information". Mimeo, Princeton University, 1992.

---

[16]For the case where $t_i^n < b_i^n$ one need only replace the roles of $\beta$ and $\gamma$.

[2] Dilip Abreu and Arunava Sen. "Subgame perfect implementation: A Necessary and Almost Sufficient Condition". *Journal of Economic Theory*, 50:285-299, 1990.

[3] A. Acquisti and H. Varian, "Conditioning Prices on Purchase History." *Marketing Science*, vol. 24, no. 3, 2005.

[4] N. Alon, M. Feldman, A. D. Procaccia and M. Tennenholtz. "Strategyproof approximation of the minimax on networks". *Math. Oper. Res.*, 35(3):513–526, 2010.

[5] P. Arabie, J. D. Carroll, W. DeSarbo, and J. Wind, "Overlapping clustering: A new method for product positioning". *Journal of Marketing Research*, vol. 18, no. 3, 1981.

[6] S. Banerjee, H. Konishi and T. Sonmez. "Core in a Simple Coalition Formation Game". *Social Choice and Welfare*, 18(1):135–153, 2001.

[7] F.M. Bass, D.J. Tigert, R.T. Lonsdale "Market segmentation: Group versus individual behavior". *Journal of Marketing Research*, V.5, No. 3, pp. 264-270, 1968.

[8] D. Bergemann and S. Morris. "Robust virtual implementation". *Theoretical Economics*, Vol. 4, No. 1, pp. 45-88, 2009.

[9] F. Bloch and E. Diamantoudi. "Noncooperative formation of coalitions in hedonic games". *International Journal of Game Theory*, 40(2):263-280, 2010.

[10] A. Bogomolnaia and M.O. Jackson. "The stability of hedonic coalition structures". *Games and Economic Behavior*, 38(2):201-230, 2002.

[11] Kamalika Chaudhuri, Anand D. Sarwate, and Kaushik Sinha. "A near-optimal algorithm for differentially private principal components". *Journal of Machine Learning Research*, 14(1): 2905-2943, 2013.

[12] M. Charikar, S. Guha, E. Tardos and D. Shmoys. A "Constant factor approximation algorithm for the k-median problem". *ACM symposium on Symposium on theory of computing* (STOC), pages 1–10. ACM, 1999.

[13] P.M. Chisnall, Marketing: a behavioural analysis, McGraw-Hill (UK), 1985

[14] R. Cummings, K. Ligett, M. Pai, and A. Roth. "The strange case of privacy in equilibrium models". *ACM Conference on Economics and Computation* (EC), 2016.

[15] J. Duggan. "Virtual Bayesian implementation." *Econometrica*, 65:1175–1199, 1997.

[16] C. Dwork, F. McSherry, K. Nissim, and A. Smith. "Calibrating noise to sensitivity in private data analysis". In Shai Halevi and Tal Rabin, editors, *TCC*, volume 3876 of *Lecture Notes in Computer Science*, pages 265–284. Springer, 2006.

[17] C. Dwork, M. Naor, O. Reingold, G. N. Rothblum, and S. Vadhan. "On the complexity of differentially private data release: Efficient algorithms and hardness results". *ACM Symposium on Theory of Computing* (STOC), pages 381-390, New York, NY, USA, 2009. ACM.

[18] M. Feldman, L. Lewin-Eytan and J. Naor. "Hedonic clustering games". *Symposium on Parallelism in Algorithms and Architectures* (SPAA), pages 267–276, 2012.

[19] D. Fotakis and C. Tzamos. "Winner-imposing strategyproof mechanisms for multiple facility location games". *Conference on Web and Internet Economics* (WINE), pages 234–245, 2010.

[20] S. Goyat. "The basis of market segmentation: a critical review of literature" *European Journal of Business and Management*, Vol 3, No.9, 2011.

[21] J.E. Hartley (1996): "Retrospectives: The origins of the representative agent", In *Journal of Economic Perspectives* 10: 169-177.

[22] O. Heffetz and K. Ligett "Privacy and Data-Based Research", *Journal of Economic Perspectives*, 28,2: 7598. 2014.

[23] D.S. Hochbaum and D.B. Shmoys. "A best possible Heuristic for the k-center problem". *Mathematics of Operations Research*, 10:180–184, 1985.

[24] H. Hotelling "Stability in Competition". *The Economic Journal* 39 (153): 4157, 1929.

[25] Z. Huang and S. Kannan. "The Exponential Mechanism for Social Welfare: Private, Truthful, and Nearly Optimal." *Symposium on Foundations of Computer Science* (FOCS), pages 140-149, 2012.

[26] IAB report "IAB Internet Advertising Revenue Report conducted by PriceWaterhouseCoopers (PWC)." `http://www.iab.net/media/file/IAB_Internet_Advertising_Revenue_FY_2014.pdf`

[27] J.P. Johnson and D.P. Myatt "On the Simple Economics of Advertising, Marketing, and Product Design". *American Economic Review*, Vol. 96, No. 3, pp. 756-784. 2006.

[28] M. Kearns, M. Pai, A. Roth, and J. Ullman, "Mechanism Design in Large Games: Incentives and Privacy." *Innovations in Theoretical Computer Science* (ITCS), 2014.

[29] P. Kotler, G. Armstrong, J. Saunders, and V. Wong, Principles of Marketing (3rd European ed.). London: Prentice-Hall.

[30] K. Lancaster "Socially Optimal Product Differentiation". *American Economic Review*, Vol. 65, No. 4, pp. 567-585. 1975.

[31] P. Lu, X. Sun, Y. Wang, and Z.A. Zhu. "Asymptotically optimal strategy-proof mechanisms for two-facility games". *ACM Conference on Electronic Commerce* (EC), pages 315–324, 2010.

[32] H. Matsushima. "A New Approach to the Implementation Problem." *Journal of Economic Theory*, 45:128–144, 1988.

[33] F. McSherry and K. Talwar. "Mechanism Design via Differential Privacy." *Symposium on Foundations of Computer Science* (FOCS), pages 94–103, 2007.

[34] K. Nissim, C. Orlandi and R. Smorodinsky. "Privacy-aware mechanism design". *ACM Conference on Electronic Commerce* (EC), pages 774–789, 2012.

[35] K. Nissim, R. Smorodinsky and M. Tennenholtz. "Approximately optimal mechanism design via differential privacy". *Innovations in Theoretical Computer Science* (ITCS), pages 203–213, 2012.

[36] M. Pai and A. Roth "Privacy and Mechanism Design", SigEcom Exchanges 12 (1), 8–29. 2014.

[37] A. Procaccia and M. Tennenholtz. Approximate*ACM Conference on Electronic Commerce* (EC), 2009.

[38] J. Schummer and R.V. Vohra. "Mechanism design without money". In E. Tardos V. Vazirani N. Nisan, T. Roughgarden, editor, *Algorithmic Game Theory*, volume 2, pages 110–130. 2007.

[39] R. Serrano and R. Vohra. "Some Limitations of Virtual Bayesian Implementation." *Econometrica*, 69:785-792, 2001.

[40] R. Serrano and R. Vohra. "A Characterization of Virtual Bayesian Implementation." *Games and Economic Behavior*, 50:312-331, 2005.

[41] W. Smith, "Product differentiation and market segmentation as alternative marketing strategies". *Journal of Marketing*, 21, 3-8. 1956.

[42] L. A. Stole. "Price Discrimination and Competition." In: Mark Armstrong and Robert Porter (Eds.), Handbook of Industrial Organization, vol. 3, Chapter 34, 2221–2292, North-Holland. 2007.

[43] J. Ullman. "Answering $n^{2+o(1)}$ counting queries with differential privacy is hard". *ACM symposium on Symposium on theory of computing* (STOC), pages 361-370. 2013.

[44] J. Ullman and S. Vadhan. "PCPs and the hardness of generating private synthetic data". *Theory of Cryptography*, pages 400-416. Springer, 2011.

[45] H. Varian. "Price discrimination. In: Schmalensee, R., Willig, R.D. (Eds.), Handbook of Industrial Organization, vol. 1., Chapter 10, 597–564, North-Holland. 1989.

[46] J. Miguel Villas-Boas "Price Cycles in Markets with Customer Recognition." *The RAND Journal of Economics*, Vol. 35, No. 3 (Autumn, 2004), pp. 486-501

[47] M. Wedel and W.A. Kamakura. Market Segmentation: Conceptualand Methodological Foundations. Kluwer, Dordrecht, The Netherlands. 2000.

[48] Y. Wind, "Issues and Advances in Segmentation Research." *Journal of Marketing Research*, 317-337, Aug 1978